\documentclass[%
 reprint,
 amsmath,amssymb,
 aps
]{revtex4-2}

\usepackage{graphicx}% Include figure files
\usepackage{dcolumn}% Align table columns on decimal point
\usepackage{bm}% bold math
\usepackage{makecell}
\usepackage[normalem]{ulem}
\usepackage{xcolor}
\usepackage{newtxtext}
\usepackage{lmodern,pdftexcmds}
\usepackage{xstring}
\usepackage{mathrsfs}
\usepackage{soul}
\usepackage{siunitx}
\usepackage{booktabs}
\usepackage{multirow}

\DeclareMathAlphabet{\mathbfsf}{\encodingdefault}{\sfdefault}{bx}{sl}
\renewcommand{\Vec}[1]{\bm{#1}}
\def\Tens#1{\IfSubStr{ABCDEFGHIJKLMNOPQRSTUVWXYZabcdefghijklmnopqrstuvwxyz}{#1}{\mathbfsf{#1}}{\bm{#1}}}

\begin{document}

\preprint{APS/123-QED}

\title{Response of magnetic particle to rotating magnetic field in viscoelastic fluid}

\author{Han \surname{Gao}$^{1,*}$}
\author{Zhiyuan \surname{Zhao}$^{2,3}$}\thanks{These authors contributed equally to this work.}
\author{Masao \surname{Doi}$^{2,3,\dag}$}
\author{Ye \surname{Xu}$^{1,}$}

\email{doi.masao@a.mbox.nagoya-u.ac.jp}
\email{ye.xu@buaa.edu.cn}

\affiliation{
$^1$School of Mechanical Engineering and Automatioen, Beihang University, 
Beijing 100191, China}
\affiliation{
$^2$Wenzhou Institute, University of Chinese Academy of Sciences, Wenzhou, Zhejiang 325000, China}
\affiliation{
$^3$Oujiang Laboratory (Zhejiang Lab for Regenerative Medicine, Vision and Brain Health), Wenzhou, Zhejiang 325000, China}

\date{\today}

\begin{abstract}

The rotational dynamics of a freely suspended ferromagnetic particle in viscoelastic fluid subjected to a rotating magnetic field is studied by experiments and theory. 
Our result reveals that when the characteristic relaxation time of the fluid is much smaller than the inverse critical field frequency, the particle’s rotation behavior aligns with that in Newtonian fluids. 
Increasing the relaxation time enhances the time-averaged rotation frequency of the particle that undergo asynchronous rotation. 
Moreover, the critical frequency is shown to scale linearly with the magnetic field intensity and inversely with the fluid’s zero-shear viscosity.
Our work is expected to guide precise manipulation of ferromagnetic particles in biomedical systems where viscoelastic environments dominate. 

\end{abstract}

% \keywords{Suggested keywords}%Use showkeys class option if keyword display desired

\maketitle

%%%%%%%%%%%%%%%%%%%%%%%%%%%%%%%%%%%%%%%%%%%%%%%%%%%%%%%%%%%%%%%%%%%%%%%%%%%%
%%%%%%%%%%%%%%%%%%%%%%%%%%%%%%%%%%%%%%%%%%%%%%%%%%%%%%%%%%%%%%%%%%%%%%%%%%%%

\section{\label{sec:level1}INTRODUCTION}

%%%%%%%%%%%%%%%%%%%%%%%%%%%%%%%%%%%%%%%%%%%%%%%%%%%%%%%%%%%%%%%%%%%%%%%%%%%%

Magnetic particles have shown great potential in biomedical technologies because of their sensitive yet controllable responses to externally-imposed magnetic fields~\cite{spatafora2021hierarchical,rivera2021emerging} 
Their applications span diverse domains including targeted drug delivery~\cite{liu2019review}, magnetic hyperthermia therapy~\cite{hedayatnasab2017review,shasha2021nonequilibrium}, advanced imaging modalities~\cite{tay2024magnetic,velazquez2025advances}, and microrobots~\cite{yu2019active,xie2019reconfigurable,yang2021motion}.
One intriguing and functional case is that in rotating magnetic fields, an isolated magnetic particle can spin following the field~\cite{mcnaughton2007physiochemical,janssen2009controlled} and even generate net translation when a boundary surface is nearby~\cite{goldman1967slow}. 
For a collection of the rotating magnetic particles, they can self-organize into reconfigurable swarms in various patterns~\cite{yu2019active,yang2021motion}, which possess abilities like moving upstream against flow~\cite{wang2021ultrasound,wang2023reconfigurable}, grasping targeted objects~\cite{sun2021swarming}, adapting to local environment~\cite{yu2021adaptive,wang2021endoscopy}, and targeted therapy~\cite{tasci2017enhanced,mellal2021modeling}.

%%%%%%%%%%%%%%%%%%%%%%%%%%%%%%%%%%%%%%%%%%%%%%%%%%%%%%%%%%%%%%%%%%%%%%%%%%%%

In order to precisely manipulate the particle motion and to optimize the performance of relevant applications, it is crucial to understand the frequency response of individual magnetic particles to imposed rotating magnetic fields. 
Prior works~\cite{mcnaughton2007physiochemical} have reported that in unbounded Newtonian fluids, the time-averaged rotation frequency $\langle \omega \rangle$ of a spherical magnetic particle is given by
\begin{equation}
    \langle \omega \rangle = 
    \begin{cases}
        \Omega, &\Omega \le \Omega_\mathrm{c} \\
        \Omega - \sqrt{ \Omega^2 - \Omega^2_\mathrm{c}} , 
        &\Omega > \Omega_\mathrm{c}
    \end{cases}
    \label{eq:frequency_newtonian}
\end{equation}
where $\Omega$ denotes the driving field frequency, the critical field frequency $\Omega_\mathrm{c} \equiv m B / \zeta_\mathrm{r}$, $\zeta_\mathrm{r}$ represents the rotational drag coefficient that is proportional to the fluid viscosity, and $m$ and $B$ are the intensities of the magnetic dipole moment and the field, respectively. 
Equation~(\ref{eq:frequency_newtonian}) shows that the particle always synchronously rotates with the field when $\Omega \le \Omega_\mathrm{c}$. 
For $\Omega > \Omega_\mathrm{c}$, the particle's angular velocity starts to change periodically in time, with the average value decreasing as the field frequency increases.

%%%%%%%%%%%%%%%%%%%%%%%%%%%%%%%%%%%%%%%%%%%%%%%%%%%%%%%%%%%%%%%%%%%%%%%%%%%%

The prediction of Eq.~(\ref{eq:frequency_newtonian}) enables unique  advantages of magnetic particles in biosensor fabrication and microrobot manipulation~\cite{sinn2011asynchronous,koo2009nanoparticle,xie2020bioinspired}. 
However, due to the existence of high molecular weight polymers ~\cite{fung2013biomechanics,hwang1969rheological,errill1969rheology},
biological fluids are usually non-Newtonian and exhibit viscoelastic behavior combining viscous dissipation and elastic response to imposed deformation. 
Such a rheological response fundamentally alters mechanical interactions at the particle-fluid interface, which potentially modify the particle dynamics predicted by conventional Newtonian models.
A review of literature suggests that the understanding of relationship between fluid viscoelasticity and magnetic particle dynamics under rotating fields is still lacking. 

%%%%%%%%%%%%%%%%%%%%%%%%%%%%%%%%%%%%%%%%%%%%%%%%%%%%%%%%%%%%%%%%%%%%%%%%%%%%

In this work, we studied the rotational motion of isolated ferromagnetic sphere that was suspended in viscoelastic fluids and driven by rotating magnetic fields. 
The particle’s average rotation frequency was experimentally measured in fluids of various compositions and viscoelastic parameters. 
The result was then explained and reproduced through a theoretical study that is based on the framework of linear viscoelasticity. 
In the analysis, we also investigated the dependence of the average rotation frequency on the fluid viscosity and relaxation time. 
At last, by comparing our results with Eq.~(\ref{eq:frequency_newtonian}), differences of the particle's rotational responses in Newtonian and viscoelastic fluids were discussed.

%%%%%%%%%%%%%%%%%%%%%%%%%%%%%%%%%%%%%%%%%%%%%%%%%%%%%%%%%%%%%%%%%%%%%%%%%%%%
%%%%%%%%%%%%%%%%%%%%%%%%%%%%%%%%%%%%%%%%%%%%%%%%%%%%%%%%%%%%%%%%%%%%%%%%%%%%

\section{Experimental Methods}

We chose two distinct types of viscoelastic fluid with each in various concentrations. 
One was prepared by mixing polyacrylamide powder (PAAM, Sigma Aldrich, 92560, 5--6MDa, non-ionic) with deionized water at ratios of $\rho = $0.05\%, 0.15\%, 0.2\%, 0.25\%, and 0.3\% w/v. 
The obtained PAAM solutions were heated at 30$^\circ$C and agitated by magnetic stirrers at 200 rpm for overnight before using~\cite{rogowski2021symmetry}. 
The other type of viscoelastic fluid was produced by mixing equimolar ratio of cetyltrimethylammonium bromide (CTAB, Sigma Aldrich, purity $>$ 99$\%$) and sodium salicylate (NaSal, Sigma Aldrich, purity $>$ 99.5$\%$) with deionized water. 
The obtained CTAB/NaSal solutions, in concentrations of 2 mM and 3 mM, were agitated by magnetic stirrer for a day and then left to set for another day before using. 

%%%%%%%%%%%%%%%%%%%%%%%%%%%%%%%%%%%%%%%%%%%%%%%%%%%%%%%%%%%%%%%%%%%%%%%%%%%%

Rheological characterization of the PAAM solutions and CTAB/NaSal solutions was conducted through complementary bulk and microrheological analyses. 
Steady-state viscometric measurements were performed using a stress-controlled rheometer (TA Instruments Discovery Hybrid Rheometer DHR-2) equipped with a cone-plate geometry (40 mm diameter, 4$^\circ$ cone angle). 
For each solution, shear rate sweeps spanning from 1 to 100 (1/s) were executed for three times, with 50-second signal stabilization interval per data point.
The zero-shear viscosity was subsequently determined by fitting the steady shear viscosity data (see Fig.\,\ref{fig:S1} in Appendix~\ref{app:A}) to the Carreau-Yasuda model~\cite{MAHMOOD20201785}.

%%%%%%%%%%%%%%%%%%%%%%%%%%%%%%%%%%%%%%%%%%%%%%%%%%%%%%%%%%%%%%%%%%%%%%%%%%%%

In the particle tracking microrheology~\cite{m2020multiple}, high-speed video microscopy was employed to capture Brownian trajectories of fluorescent particles (Thermo Fisher, 0.11um, 580nm/605nm) dispersed in different solution samples.
The mean square displacements (MSDs) of the fluorescent particles were then estimated according to~\cite{kim2021measuring}. 
We performed such measurements in different locations of the solutions and repeated for twice with 10 data sets for each sample.
Fig.\,\ref{fig:S2} (Appendix~\ref{app:A}) shows the representative time-dependent MSDs in the PAAM solutions, where accordance of the curves suggests the homogeneous distribution of polymers. 
Based on the MSD data, we computed the elastic moduli of the solutions and then extracted the corresponding relaxation time by fitting such moduli to a single-mode Maxwell model~\cite{PhysRevLett.85.1774}.

%%%%%%%%%%%%%%%%%%%%%%%%%%%%%%%%%%%%%%%%%%%%%%%%%%%%%%%%%%%%%%%%%%%%%%%%%%%%

% Figure 1
\begin{figure}[tb]
	\includegraphics[width=0.48\textwidth]{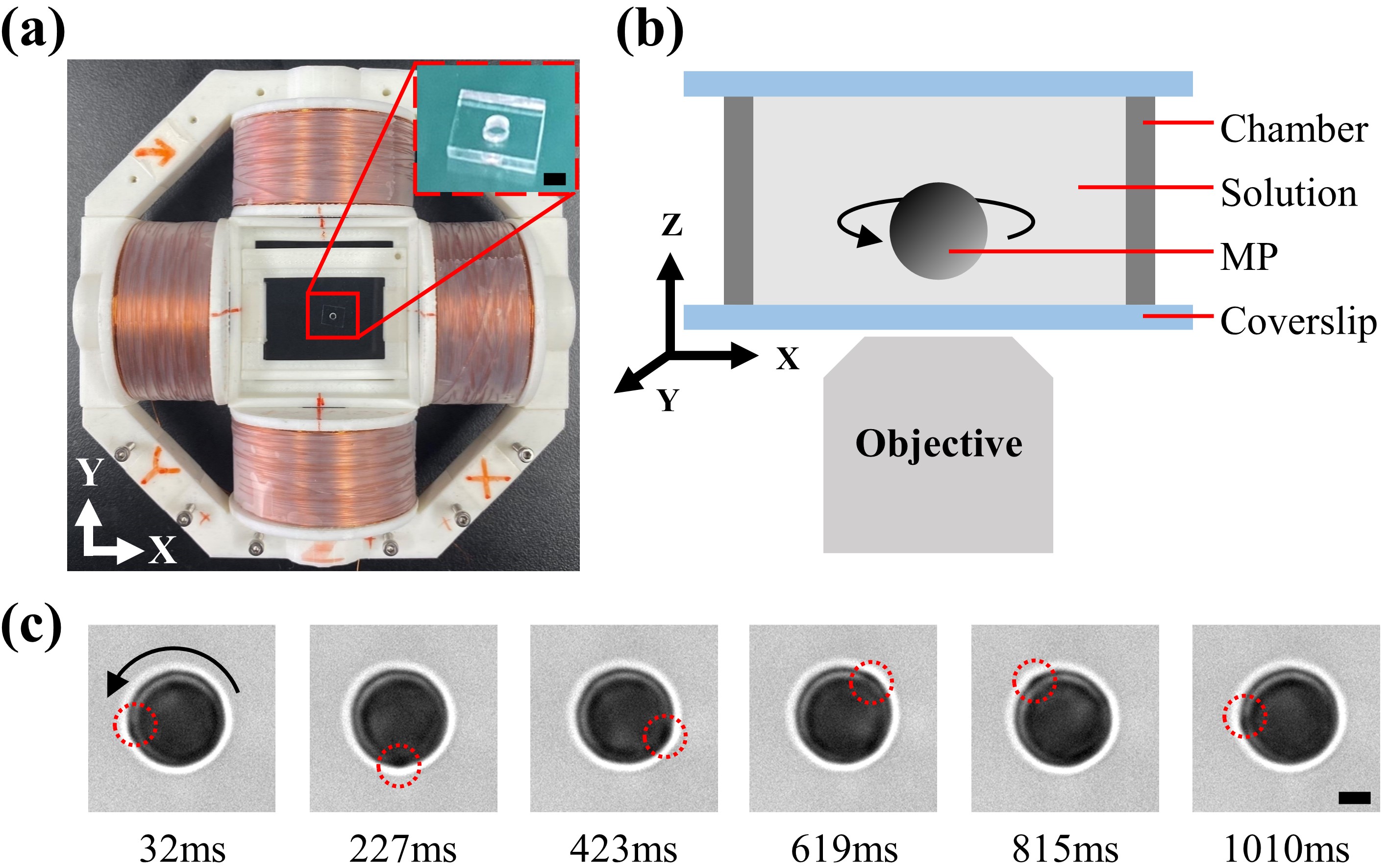}
	\caption{\label{fig:1}
    (a) Top view of experimental setup. The customized magnetic field generator is composed of two pairs of orthogonal solenoids and can generate uniform rotating magnetic fields (in the $x$-$y$ plane) in the center region. Insert shows a PDMS cube with a cylindrical sample chamber at the center. The scale bar indicates 1 mm.
    (b) Schematic side view of the experimental setup. The sample chamber, which is filled with dilute magnetic particle suspension and sealed by two parallel coverslips, is set on the stage of an inverted microscope. 
    (c) Representative snapshots of a rotating magnetic particle that is suspended in the PAAM solution of concentration 0.15\%. The driving magnetic field is rotating in the counterclockwise direction with the amplitude 0.71 mT and frequency 1 Hz. The red dashed circles highlight the particle characteristic pattern for rotation tracing. The scale bar corresponds to 1 $\mu$m. 
    }
\end{figure}

%%%%%%%%%%%%%%%%%%%%%%%%%%%%%%%%%%%%%%%%%%%%%%%%%%%%%%%%%%%%%%%%%%%%%%%%%%%%

The experiment samples were prepared by diluting 2.5 $\si{\micro \litre}$ of commercial ferromagnetic particle suspension (Tianjin BaseLine, Affimag $\gamma$-Fe$_2$O$_3$, diameter 3 to 4 $\si{\micro \meter}$, 1\% w/v, --NH$_2$) into 1 $\si{\milli \litre}$ of each of the PAAM solutions, CTAB/NaSal solutions, and deionized water. 
To ensure uniform dispersion, the samples were also subjected to a high-speed vortex mixer (1000 rpm) for 30 min. 
It is noted that the sample with the deionized water is only for contrast, where saturated sodium dodecyl sulfate (SDS) solution (at a ratio of 1\% v/v) was added to prevent particles from sticking to glass substrates~\cite{martinez2018emergent}. 

%%%%%%%%%%%%%%%%%%%%%%%%%%%%%%%%%%%%%%%%%%%%%%%%%%%%%%%%%%%%%%%%%%%%%%%%%%%%

We loaded 5 $\mu$L of above samples into individual polydimethylsiloxane (PDMS) chambers (1 mm diameter $\times$ 1 mm depth) with glass substrates. 
Immediately after loading, the chambers were sealed with coverslips in order to inhibit evaporation and internal flows of the samples. 
Then the assembled chambers were centrally positioned within a customized magnetic field generator (see Fig.\,\ref{fig:1}(a)), which was mounted to an inverted microscope (Nikon Ti-2) with a 100$\times$ oil-immersion objective (Nikon Plan Apo $\lambda$, 1.45 N.A.). 
The side view of the setup is schematically illustrated in Fig.\,\ref{fig:1}(b).

%%%%%%%%%%%%%%%%%%%%%%%%%%%%%%%%%%%%%%%%%%%%%%%%%%%%%%%%%%%%%%%%%%%%%%%%%%%%

The rotating magnetic field was generated using two pairs of homemade orthogonal solenoids, with the field amplitude and frequency regulated by a function generator (DG1062Z, RIGOL) and a power amplifier (ATA-2022B, Agitek), respectively. 
The percent variation of the field intensity within 4 mm $\times$ 4 mm at the center of the working plane, where the PDMS chamber commonly locates, is lower than 6\%.
The expression of such a time-dependent magnetic field is $\Vec{B}(t) = B_0 [\cos (\Omega t) \Vec{e}_x + \sin (\Omega t) \Vec{e}_y]$, where $B_0$ denotes the field intensity, $\Omega$ the angular frequency, $t$ the time, and $\Vec{e}_x$ and $\Vec{e}_y$ the unit vectors in the $x$ and $y$ directions, respectively. 
 
%%%%%%%%%%%%%%%%%%%%%%%%%%%%%%%%%%%%%%%%%%%%%%%%%%%%%%%%%%%%%%%%%%%%%%%%%%%%

The particle rotation was captured using a CMOS camera (Photometrics Prime BSI) with a pixel resolution of 500 $\times$ 500 at 250 frames/s. 
All the samples were confirmed to be in the dilute regime, justifying the negligibility of inter-particle interactions. 
Additionally, each particle possesses own intrinsic characteristic patterns that can specify its orientation, as shown in Fig.\,\ref{fig:1}(c) and MOVIE\,S1 (Supplemental Information). 
This makes us available to measure the particle's rotational displacement within certain time and then estimate the corresponding time-averaged rotation frequency over 5 revolutions~\cite{janssen2009controlled}.

%%%%%%%%%%%%%%%%%%%%%%%%%%%%%%%%%%%%%%%%%%%%%%%%%%%%%%%%%%%%%%%%%%%%%%%%%%%%
%%%%%%%%%%%%%%%%%%%%%%%%%%%%%%%%%%%%%%%%%%%%%%%%%%%%%%%%%%%%%%%%%%%%%%%%%%%%

\section{RESULTS AND DISCUSSION}

%%%%%%%%%%%%%%%%%%%%%%%%%%%%%%%%%%%%%%%%%%%%%%%%%%%%%%%%%%%%%%%%%%%%%%%%%%%%

The viscoelastic parameters of the PAAM solutions and CTAB/NaSal solutions are detailed in TABLEs\,I and II, respectively. 
It is observed that both zero-shear viscosity ($\eta_\mathrm{tot}$) and characteristic relaxation time ($\tau$) monotonically increases when the polymer solution becomes concentrated. 
We did not consider concentrations larger than 0.3$\%$ w/v for the PAAM solution due to its saturation limit, nor concentrations larger than 3 mM for the CTAB/NaSal solution, because under such conditions the magnetic particle hardly rotates. 

%%%%%%%%%%%%%%%%%%%%%%%%%%%%%%%%%%%%%%%%%%%%%%%%%%%%%%%%%%%%%%%%%%%%%%%%%%%%

% Table 1
\begin{table}[tb]
    \caption{
    \label{tab:1} 
    Rheological parameters of the PAAM solution in various w/v ratios, where $\eta_\mathrm{tot}$ and $\tau$ denote the zero-shear viscosity and relaxation time of the solution, respectively, and $\Omega^\ast_\mathrm{c,exp}$ represents the experimental critical field frequency.
    }
    \begin{ruledtabular}
    \begin{tabular}{cccccc}
    \makecell{$\rho$ \\ (\% w/v)} 
        & \makecell{$\eta_\mathrm{tot}$ \\ (mPa s)} 
        & \makecell{$\tau$ \\ (ms)} 
        & \makecell{$\Omega^\ast_\mathrm{c,exp}$ \\ (Hz)}
        & $\tau \Omega^\ast_\mathrm{c,exp}$ \\ 
    \midrule
    0.05 & 2.03 & 13.8 & 8.4 & 0.1159 \\
    0.15 & 3.49 & 22.6 & 4.1 & 0.0927 \\
    0.20 & 4.78 & 25.7 & 2.6 & 0.0668 \\
    0.25 & 5.62 & 49.0 & 1.6 & 0.0784 \\
    0.30 & 8.02 & 76.6 & 1.3 & 0.0996 \\
    \end{tabular}
    \end{ruledtabular}
\end{table}

% Table 2
\begin{table}[tb]
    \caption{
    \label{tab:2} 
    Rheological parameters of the CTAB/NaSal solution in various concentrations, where the variables are defined in the same manner of those in TABLE\,I.
    }
    \begin{ruledtabular}
    \begin{tabular}{cccccc}
    \makecell{$\rho$ \\ (mM)} 
        & \makecell{$\eta_\mathrm{tot}$ \\ (mPa s)} 
        & \makecell{$\tau$ \\ (ms)} 
        & \makecell{$\Omega^\ast_\mathrm{c,exp}$ \\ (Hz)} 
        & $\tau \Omega^\ast_\mathrm{c,exp}$ \\ 
    \midrule
    2 & 3.1 & 7.6 & 28.1 & 0.2136 \\
    3 & 115.2 & 488.6 & 1.2 & 0.5863 \\
    \end{tabular}
    \end{ruledtabular}
\end{table}

%%%%%%%%%%%%%%%%%%%%%%%%%%%%%%%%%%%%%%%%%%%%%%%%%%%%%%%%%%%%%%%%%%%%%%%%%%%%

\begin{figure}[tb]
	\centering
	\includegraphics[width=0.48\textwidth]{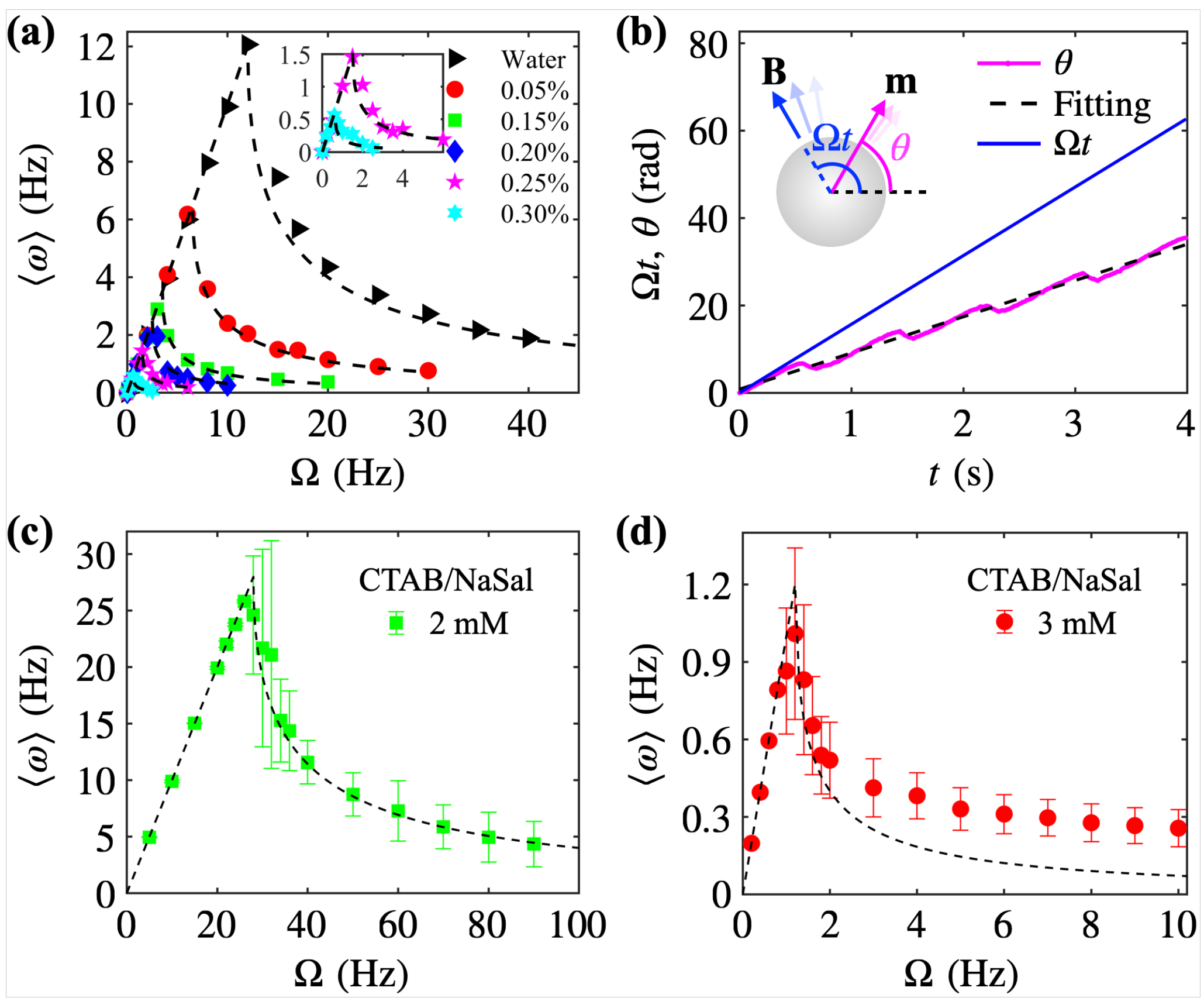}
	\caption{\label{fig:2}
    (a) Average rotation frequency $\langle \omega \rangle$ of an isolated magnetic particle as a function of field frequency $\Omega$ for solvents of water and PAAM solutions in various mass concentrations. 
    The magnetic field intensity $B_0 =$ 0.71 mT. 
    The dashed lines are fitted in terms of Eq.~(\ref{eq:frequency_newtonian}).
    (b) Time evolution of cumulated azimuthal angles of the magnetic particle ($\theta$) and the rotating magnetic field ($\Omega t$) for the case of mass concentration of the PAAM solution $\rho =$ 0.25\%, field intensity $B_0 =$ 0.71 mT, and $\Omega =$ 2.5 Hz (which is larger than the critical frequency $\Omega^\ast_\mathrm{c} =$ 1.5 Hz). 
    The dashed line is the fitting to the curve of $\theta$.
    Inset: Schematic of angles $\theta$ and $\Omega$ in terms of magnetic dipole moment $\mathbf{m}$ and magnetic field $\mathbf{B}$.
    (c) and (d) shows the dependence of $\langle \omega \rangle$ on $\Omega$ for the CTAB/NaSal solution of concentrations 2 mM and 3 mM, respectively. 
    The magnetic field intensity $B_0 =$ 4 mT. 
    The dashed lines are fitted in terms of Eq.~(\ref{eq:frequency_newtonian}).
    }
\end{figure}

Figure\,\ref{fig:2}(a) presents the rotational response of a ferromagnetic particle to an imposed rotating magnetic field for surrounding fluids of deionized water and the PAAM solutions. 
In the figure, the field intensity is $B_0 =$ 0.71 mT and $\langle \omega \rangle$ represents the time-averaged rotation frequency of the particle.
One can observe that each curve possesses a distinct critical field frequency $\Omega^\ast_\mathrm{c}$, whose magnitude demonstrates an inverse dependence on the solution concentration. 
When the field frequency is below this critical value ($\Omega < \Omega^\ast_\mathrm{c}$), the particle rotates synchronously with the driving field.
However, above this threshold ($\Omega > \Omega^\ast_\mathrm{c}$), the particle rotation frequency undergoes an abrupt decline with increasing $\Omega$. 
Furthermore and surprisingly, despite variations in polymer concentration, all the data points approach the theoretical prediction of Eq.~(\ref{eq:frequency_newtonian}) (dashed curves), showing remarkable agreement with the Newtonian fluid model. 

%%%%%%%%%%%%%%%%%%%%%%%%%%%%%%%%%%%%%%%%%%%%%%%%%%%%%%%%%%%%%%%%%%%%%%%%%%%%

%%%%%%%%%%%%%%%%%%%%%%%%%%%%%%%%%%%%%%%%%%%%%%%%%%%%%%%%%%%%%%%%%%%%%%%%%%%%

Prior work~\cite{mcnaughton2007physiochemical,janssen2009controlled} has explained that the nonlinear response of the particle rotation frequency in Newtonian fluids arises from phase lagging and then opposite rotation of the magnetic particle. 
Such nonlinear dynamics also exist in PAAM solutions.
Figure\,\ref{fig:2}(b) shows the time evolution of azimuthal angles of the particle ($\theta$) and rotating field ($\Omega t$) for representative condition of $\rho =$ 0.25\%, $B_0 =$ 0.71 mT, and $\Omega$ = 2.5 Hz (which is large than $\Omega^\ast_\mathrm{c}$ = 1.5 Hz). 
In contrast with the field orientation, $\theta$ evolves with sawtooth oscillations, reflecting `wobbling' rotation of the particle (see Movie\,S2 in Supplemental Information).
By fitting the slopes of such sawtooth curves, one can obtain the value of $\langle \omega \rangle$ in the asynchronous regime.

%%%%%%%%%%%%%%%%%%%%%%%%%%%%%%%%%%%%%%%%%%%%%%%%%%%%%%%%%%%%%%%%%%%%%%%%%%%%

Figures\,\ref{fig:2}\,(c) and (d) present the relation between $\langle \omega \rangle$ and $\Omega$ for the CTAB/NaSal solutions of concentrations 2mM and 3mM, respectively, and $B_0 =$ 4 mT. 
The particle's frequency response is also featured with linear synchronization at low frequencies and nonlinear decoupling in the high-frequency regime.
However, when the micellar concentration increases, $\langle \omega \rangle$ in the asynchronous regime enhances and diverges from the prediction of Eq.~(\ref{eq:frequency_newtonian}) (see Fig.\,\ref{fig:2}(d)). 

%%%%%%%%%%%%%%%%%%%%%%%%%%%%%%%%%%%%%%%%%%%%%%%%%%%%%%%%%%%%%%%%%%%%%%%%%%%%

In order to elucidate the influence of the fluid viscoelasticity on the particle rotation frequency yet critical field frequency, an analytical study was conducted. 
We considered that a spherical particle, embedded with fixed magnetic dipole moment $\Vec{m}$, is subjected to a rotating magnetic field of intensity $B_0$ and frequency $\Omega$. 
The magnetic torque on the particle is given by
\begin{equation}
    T_{\mathrm{m}}  = m B_0 \sin (\Omega t - \theta),
    \label{eq:magnetic_torque}
\end{equation}
where $m$ and $\theta$ denote the magnitude and azimuthal angle of the magnetic dipole moment, respectively. 
%%%%%%%%%%%%%%%%%%%%%%%%%%%%%%%%%%%%%%%%%%%%%%%%%%%%%%%%%%%%%%%%%%%%%%%%%%%%

If the particle is placed in Newtonian fluid, the magnetic torque is balanced by the viscous torque exerted by the fluid. 
The time evolution equation of $\theta$ reads
\begin{equation}
	\zeta_\mathrm{r} \dot{\theta} = m B_0 \sin (\Omega t - \theta),
    \label{eqn:D1}
\end{equation}
where $\dot \theta$ represents the angular velocity of the particle, which is related to the particle rotation frequency by $\omega = \dot \theta / (2 \pi)$, the rotational friction constant $\zeta_r = 8 \pi a^3 \eta_\mathrm{s}$, and $\eta_\mathrm{s}$ and $a$ stand for the solvent viscosity and particle radius, respectively. 
The inertia of such a system can be ignored because of the small particle size ($a \approx$ 1.5 $\mu$m).

%%%%%%%%%%%%%%%%%%%%%%%%%%%%%%%%%%%%%%%%%%%%%%%%%%%%%%%%%%%%%%%%%%%%%%%%%%%%

Equation~(\ref{eqn:D1}) is only valid for Newtonian fluid where the viscous toque at time $t$ is determined by $\dot \theta(t)$. 
However, when the particle is placed in a polymer solution, the viscous torque additionally depends on $\dot \theta$ in the past. 
In the linear viscoelastic regime, the torque balance becomes
\begin{equation}
	8 \pi a^3 \left[ \eta_\mathrm{s} \dot{\theta} 
        + \int_{-\infty}^t dt' G(t-t') \dot{\theta}(t') \right ]
	= m B_0 \sin (\Omega t - \theta) 
    \label{eqn:D2}
\end{equation}
with the relaxation modulus of the polymer solution
\begin{equation}
	G(t)= \sum_i G_i e^{-t/\tau_i}.
    \label{eqn:D3}
\end{equation}
In Eq.~(\ref{eqn:D3}), $G_i$ and $\tau_i$ are material constants, representing respectively the elastic modulus and the relaxation time that are associated with the $i$-th relaxation mode of the polymer solution. 
Then our aim is to analyze the solution of the integral-differential equation Eq.~(\ref{eqn:D2}).

%%%%%%%%%%%%%%%%%%%%%%%%%%%%%%%%%%%%%%%%%%%%%%%%%%%%%%%%%%%%%%%%%%%%%%%%%%%%

We rewrote Eq.~(\ref {eqn:D2}) in a dimensionless form of
\begin{equation}
	 \dot{\theta} 
	+ \frac{1}{\eta_\mathrm{s}}\int_{-\infty}^t dt' G(t-t') \dot{\theta}(t') 
	= \Omega_\mathrm{c} \sin (\Omega t - \theta).
    \label{eqn:D3a}
\end{equation}
It can be transformed to a set of ordinary differential equations:
\begin{equation}
    \dot \theta 
    + 
    \sum_i \frac{\eta_{\mathrm{p},i}}{\eta_\mathrm{s} \tau_i} \theta_{\mathrm{e},i} 
    = \Omega_\mathrm{c} \sin(\Omega t - \theta),
    \label{eqn:D5}
\end{equation}
where $\eta_{\mathrm{p},i} = G_i \tau_i$ denotes the polymer viscosity associated with the $i$-th relaxation mode.
The variable $\theta_{\mathrm{e},i}$, which quantifies the particle rotation coupled to elastic relaxation governed by $G_i$, is given by
\begin{equation}
    \theta_{\mathrm{e},i}
    =
    \int_{-\infty}^t dt'e^{-(t-t')/\tau_i} 
    \dot{\theta}(t').
    \label{eqn:D5a}
\end{equation}
Meanwhile, $\theta_{\mathrm{e},i}$ satisfies the equation
\begin{equation}
    \dot \theta_{\mathrm{e},i}
    = 
    - \frac{\theta_{\mathrm{e},i}}{\tau_i} 
    + 
    \dot {\theta}.
    \label{eqn:D6}
\end{equation}

%%%%%%%%%%%%%%%%%%%%%%%%%%%%%%%%%%%%%%%%%%%%%%%%%%%%%%%%%%%%%%%%%%%%%%%%%%%%

By introducing the delay angle $\phi(t)= \Omega t - \theta(t)$, we rewrote Eq.~(\ref{eqn:D5}) and Eq.~(\ref{eqn:D6}) to the following set of equations:
\begin{gather}
    \Omega - \dot \phi 
    = 
    -\sum_i \frac{\eta_{\mathrm{p},i}}{\eta_\mathrm{s} \tau_i} \theta_{\mathrm{e},i}
    + 
    \Omega_\mathrm{c} \sin \phi,
    \label{eqn:D7} \\
    \dot \theta_{\mathrm{e},i}
    = 
    - \frac{\theta_{\mathrm{e},i}}{\tau_i} + \Omega - \dot {\phi}. 
    \label{eqn:D8}
\end{gather}
In order to analytically solve such two equations, we made an approximation of $\dot \theta_{\mathrm{e},i} = 0$, which indicates that $\Omega - \dot \phi$ changes little within $\tau_i$ or more physically, the elastic relaxation of the polymer solution is much faster than the particle rotation. 
Therefore, by substituting Eq.~(\ref{eqn:D8}) into Eq.~(\ref{eqn:D7}), we obtained
\begin{equation}
    \Omega - \dot \phi =  \Omega^\ast_\mathrm{c} \sin \phi
    \label{eqn:D9}
\end{equation}
with
\begin{equation}
   \Omega^\ast_\mathrm{c}  
   = 
   \frac{mB_0}{8 \pi a^3(\eta_\mathrm{s} + \sum_i \eta_{\mathrm{p},i} )}.
   \label{eqn:D10}
\end{equation}

%%%%%%%%%%%%%%%%%%%%%%%%%%%%%%%%%%%%%%%%%%%%%%%%%%%%%%%%%%%%%%%%%%%%%%%%%%%%

Equation~(\ref{eqn:D9}) is identical with Eq.~(\ref{eqn:D1}) for Newtonian fluid.
Accordingly, the analytical solution of Eq.~(\ref{eqn:D9}) is in the same manner of Eq.~(\ref{eq:frequency_newtonian}). 
This argument suggests the congruence of the $\langle \omega \rangle$ behaviors between viscoelastic and Newtonian fluids. 
However, considering that $\eta_\mathrm{tot} \equiv \eta_\mathrm{s} + \sum_i \eta_{\mathrm{p},i}$ stands for the viscosity of the polymer solution at zero frequency, $\Omega_\mathrm{c}$ in Eq.~(\ref{eq:frequency_newtonian}) needs to be replaced by $\Omega^\ast_\mathrm{c}$ . 
It is also noted that when $\Omega^\ast_\mathrm{c}$ is taken as the unit of frequency, the curves of $\langle \omega \rangle$ become overlapped with each other.

%%%%%%%%%%%%%%%%%%%%%%%%%%%%%%%%%%%%%%%%%%%%%%%%%%%%%%%%%%%%%%%%%%%%%%%%%%%%

% Figure 3
\begin{figure}[tb]
    \centering
    \includegraphics[width=0.48\textwidth]{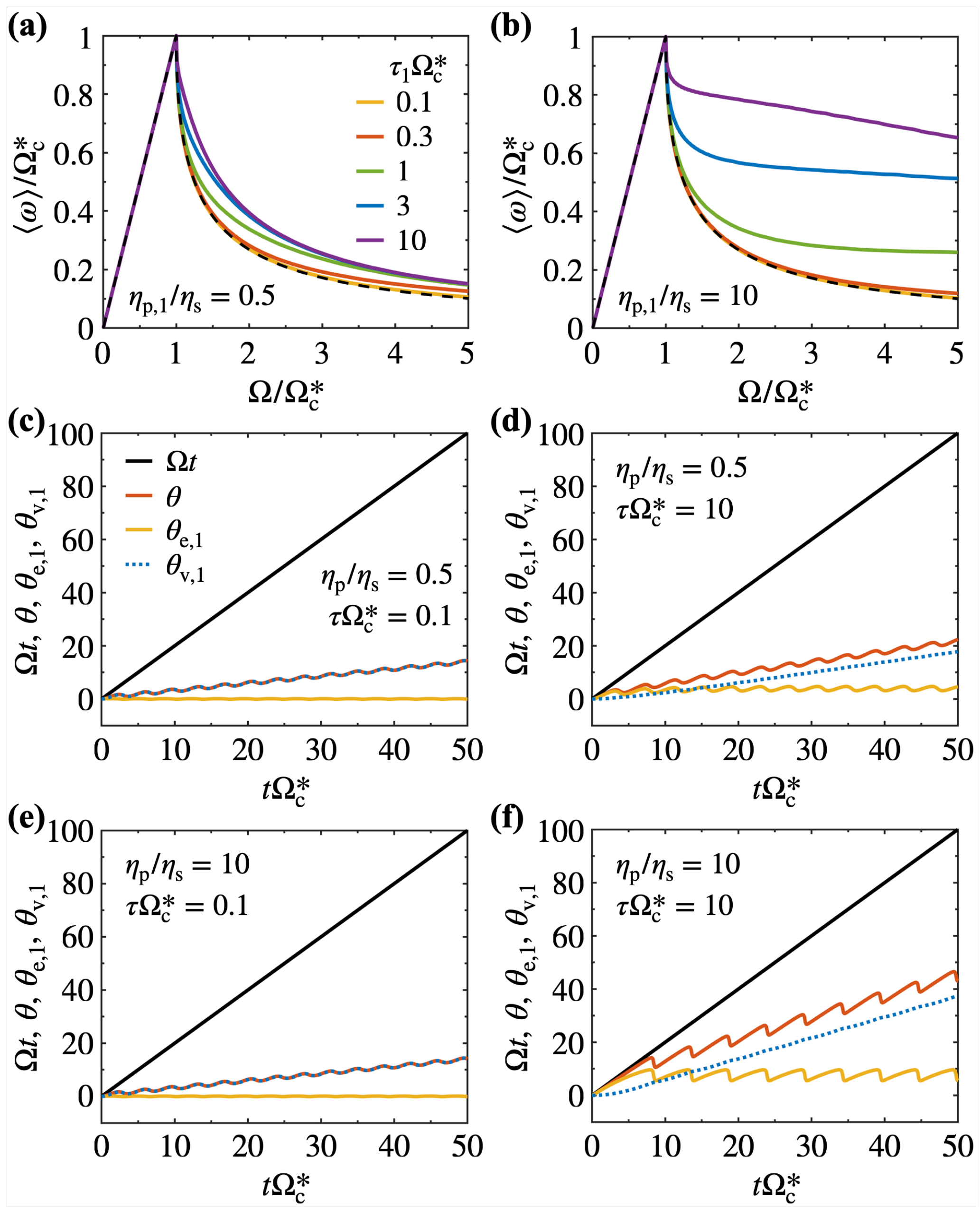}
    \caption{ \label{fig:3}
    Analytical and numerical predictions of average particle rotation frequency $\langle \omega \rangle$ as a function of frequency $\Omega$ of the rotating magnetic field, for the viscosity ratios (a) $\eta_\mathrm{p} / \eta_\mathrm{s} = 0.5 $ and (b) $10$ and for various values of single relaxation time $\tau$. 
    The black dashed lines denote the analytical prediction of Eq.~(\ref{eq:frequency_newtonian}).
    All the time-related variables are non-dimensionalized by $\Omega^\ast_\mathrm{c}$.
    Time evolution of key angular variables for (a) $\eta_\mathrm{p} / \eta_\mathrm{s} = 0.1$ and $\tau \Omega^\ast_\mathrm{c} = 0.1$, (b) $\eta_\mathrm{p} / \eta_\mathrm{s} = 0.1$ and $\tau \Omega^\ast_\mathrm{c} = 10$, (c) $\eta_\mathrm{p} / \eta_\mathrm{s} = 10$ and $\tau \Omega^\ast_\mathrm{c} = 0.1$, and (d) $\eta_\mathrm{p} / \eta_\mathrm{s} = 10$ and $\tau \Omega^\ast_\mathrm{c} = 10$. 
    Here, $\theta$ represents the cumulated angles of the magnetic particle, $\Omega t$ the rotating magnetic field, $\phi$ particle phase lag, and $\theta_\mathrm{e,1}$ and $\theta_\mathrm{v,1}$ the particle orientation components coupled to single-mode elastic and viscous relaxations, respectively.
    }
\end{figure}

%%%%%%%%%%%%%%%%%%%%%%%%%%%%%%%%%%%%%%%%%%%%%%%%%%%%%%%%%%%%%%%%%%%%%%%%%%%%

In order to examine the accuracy of the approximation, we solved Eqs.~(\ref{eqn:D7}) and (\ref{eqn:D8}) using the Runge-Kutta method and by assuming a single-mode exponential relaxation $G(t) = G_1 \exp(-t/\tau_1)$ with $\tau_1 = G_1 / \eta_\mathrm{p,1}$. 
The comparison between the numerical results and the analytical solution is shown in Figs.\,\ref{fig:3}(a) and (b) for $\eta_\mathrm{p,1}/\eta_\mathrm{s} = 0.5$ and $10$, respectively. 
It is observed that in the synchronous regime ($\Omega / \Omega^\ast_\mathrm{c} < 1$), all the numerical results are consistent with the analytical solution (dashed lines). 
This is because the two-dimensional dynamic system, governed by $\phi$ and $\theta_{\mathrm{e},1}$ (denoting $\theta_{\mathrm{e},i}$ for the single-mode exponential relaxation), achieves a steady state with $\dot \phi = 0$ and $\dot \theta_{\mathrm{e},1} = 0$.
The unique stable fixed point locates at $\phi = \arcsin (\Omega / \Omega^\ast_\mathrm{c})$ and $\theta_{\mathrm{e},1} = \tau_1 \Omega$.
Hence, tuning the viscosity and relaxation time of the viscoelastic fluids solely leads to relocation of the stable fixed point but invariance of the particle's rotation frequency. 
Such a behavior is illustrated in Fig.\,\ref{fig:S3} (Appendix~\ref{app:A}) that depicts the time evolution of key angular variables for representative condition of $\eta_\mathrm{p,1} / \eta_\mathrm{s} = 10$ and $\Omega / \Omega^\ast_\mathrm{c} = 0.5$. 

%%%%%%%%%%%%%%%%%%%%%%%%%%%%%%%%%%%%%%%%%%%%%%%%%%%%%%%%%%%%%%%%%%%%%%%%%%%%

Nonetheless, in the asynchronous regime ($\Omega / \Omega^\ast_\mathrm{c} > 1$) of Figs.\,\ref{fig:3}(a) and (b), the numerically-computed $\langle \omega \rangle$ enhances with growing $\tau$, resulting in a deviation from the analytical prediction.
Equation~(\ref{eqn:D9}) is only valid when $\tau_1 \Omega^\ast_\mathrm{c} \rightarrow 0$, i.e., the elastic effect vanishes. 
A further comparison between Figs.\,\ref{fig:3}(a) and (b) reveals that the enhancement of $\langle \omega \rangle$ becomes pronounced as the polymer viscosity increases. 
Such results suggest nontrivial influence of both the fluid viscosity and elasticity on the particle's asynchronous rotation. 

%%%%%%%%%%%%%%%%%%%%%%%%%%%%%%%%%%%%%%%%%%%%%%%%%%%%%%%%%%%%%%%%%%%%%%%%%%%%

Figures\,\ref{fig:3}(c) to (f) show the time evolution of the key angular variables for representative conditions in the asynchronous regime ($\Omega / \Omega^\ast_\mathrm{c} = 2$). 
The angle $\theta_{\mathrm{v},1} = \theta - \theta_{\mathrm{e},1}$ measures the particle rotation coupled to polymers' viscous relaxation governed by $\eta_{\mathrm{p},1}$. 
The curves of $\theta$ and $\theta_{\mathrm{v},1}$ exhibit identical time-averaged slopes and even overlap at $\tau \Omega^\ast_\mathrm{c} = 0.1$.
Essentially, the sawtooth shape of the $\theta$ curve is dominated by the viscous effect at $\tau \Omega^\ast_\mathrm{c} = 0.1$ but by elastic effect when $\tau \Omega^\ast_\mathrm{c}$ increases to $10$.
The transition of such a dominance is further illustrated in Fig.\,\ref{fig:S4} (Appendix~\ref{app:A}).
The result shows that increasing $\tau$ amplifies the sawtooth amplitude, period, and the time-averaged value of $\theta_{\mathrm{e},1}$.
Especially, the particle's forward rotation becomes more significant than the backward rotation, causing the enhanced net rotation and $\langle \omega \rangle$ value at large $\tau$. 

%%%%%%%%%%%%%%%%%%%%%%%%%%%%%%%%%%%%%%%%%%%%%%%%%%%%%%%%%%%%%%%%%%%%%%%%%%%%

% Figure 4
\begin{figure}[tb]
    \centering
    \includegraphics[width=0.48\textwidth]{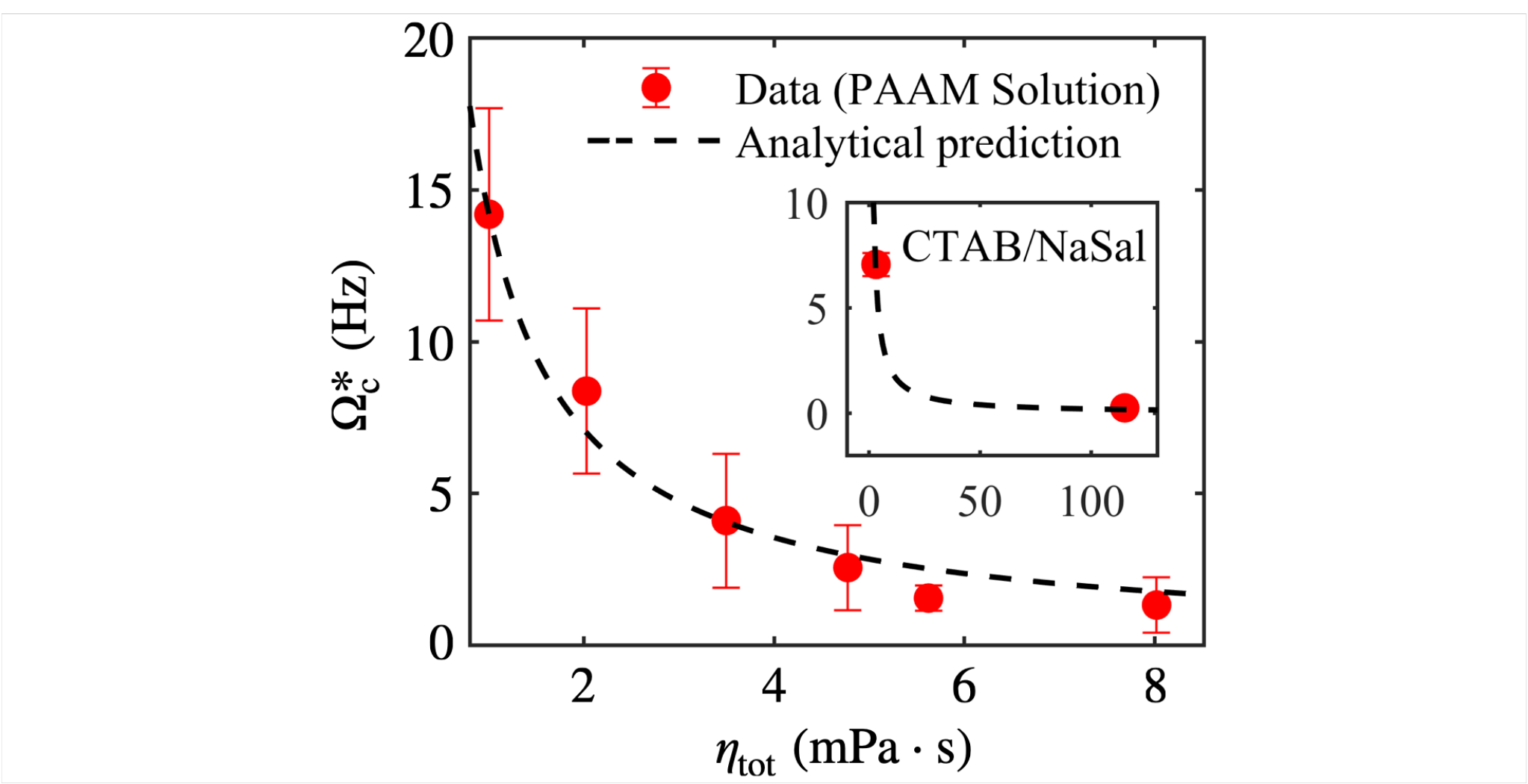}
    \caption{\label{fig:4}
    Dependence of the critical field frequency $\Omega^\ast_\mathrm{c}$, which is fitted from Fig.~\ref{fig:2}(a) according to Eq.~(\ref{eqn:D10}), on the zero shear viscosity of the PAAM solution for the magnetic field intensity $B_0 =$ 0.71 mT.
    Insert shows the same dependence but for the CTAB/NaSal solutions under $B_0 =$ 4 mT.
    }
\end{figure}

%%%%%%%%%%%%%%%%%%%%%%%%%%%%%%%%%%%%%%%%%%%%%%%%%%%%%%%%%%%%%%%%%%%%%%%%%%%%

In Fig.\,\ref{fig:4}, the analytical prediction of Eq.~(\ref{eqn:D10}) was compared with the experimental critical field frequency $\Omega^\ast_\mathrm{c,exp}$ that was fitted in terms of Eq.~(\ref{eq:frequency_newtonian}) (see Fig.~\ref{fig:2}). 
For both the PAAM solutions and the CTAB/NaSal solutions, the comparisons show with good agreement.
The values of $\Omega^\ast_\mathrm{c,exp}$ and then estimated $\tau \Omega^\ast_\mathrm{c,exp}$ for various polymer solutions are detailed in TABLEs\,I and II.
Because $\tau \Omega^\ast_\mathrm{c,exp}$ is small for all the PAAM solutions and the 2mM CTAB/NaSal solution, their experimentally-measured $\langle \omega \rangle$ values approach the prediction of Eq.~(\ref{eq:frequency_newtonian}). 
However, this is not the case for the 3mM CTAB/NaSal solution, where increased $\tau \Omega^\ast_\mathrm{c,exp}$ ($\approx 0.59$) gives rise to a certain enhancement of $\langle \omega \rangle$ in the asynchronous regime.
Such results further validate the accordance between our experimental and theoretical studies. 

%%%%%%%%%%%%%%%%%%%%%%%%%%%%%%%%%%%%%%%%%%%%%%%%%%%%%%%%%%%%%%%%%%%%%%%%%%%%
%%%%%%%%%%%%%%%%%%%%%%%%%%%%%%%%%%%%%%%%%%%%%%%%%%%%%%%%%%%%%%%%%%%%%%%%%%%%

\section{Conclusions}

By conducting experiments and theoretical analysis, we investigated the effect of fluid viscoelasticity on the rotational dynamics of an isolated ferromagnetic particle subjected to external rotating magnetic fields. 
The results demonstrate the existence of a critical field frequency $\Omega^\ast_\mathrm{c}$ as analogous to that observed in Newtonian fluids. 
When the field frequency $\Omega < \Omega^\ast_\mathrm{c}$, the particle maintains synchronous rotation with the applied field, whereas for $\Omega > \Omega^\ast_\mathrm{c}$ the particle rotation becomes asynchronous, with the time-averaged frequency decreasing as the field frequency increases.
Moreover and importantly, increasing either the zero-shear viscosity or characteristic relaxation time of the polymer solution gives rise to elevated particle rotation frequency at the asynchronous regime.
For the value of $\Omega^\ast_\mathrm{c}$, it is proportional to the applied field strength and show inverse proportionality to the fluid zero-shear viscosity.
This quantitative understanding of viscoelasticity-mediated rotational dynamics provides crucial insights for optimizing magnetic manipulation strategies in biomedical applications where viscoelastic environments are prevalent.

%%%%%%%%%%%%%%%%%%%%%%%%%%%%%%%%%%%%%%%%%%%%%%%%%%%%%%%%%%%%%%%%%%%%%%%%%%%%
%%%%%%%%%%%%%%%%%%%%%%%%%%%%%%%%%%%%%%%%%%%%%%%%%%%%%%%%%%%%%%%%%%%%%%%%%%%%

\begin{acknowledgments}

This work was supported by 
the National Nature Science Foundation of China 
(Nos.\,12072010, 11674019, and 12304247), 
the Fundamental Research Funds for the Central Universities 
(YWF-22-K-101), 
and startup funds of Wenzhou Institute, 
University of Chinese Academy of Sciences (No.\,WIUCASQD2022004)
and Oujiang Laboratory (No.\,OJQDSP2022018). 

\end{acknowledgments}

%%%%%%%%%%%%%%%%%%%%%%%%%%%%%%%%%%%%%%%%%%%%%%%%%%%%%%%%%%%%%%%%%%%%%%%%%%%%
%%%%%%%%%%%%%%%%%%%%%%%%%%%%%%%%%%%%%%%%%%%%%%%%%%%%%%%%%%%%%%%%%%%%%%%%%%%%

\appendix

%%%%%%%%%%%%%%%%%%%%%%%%%%%%%%%%%%%%%%%%%%%%%%%%%%%%%%%%%%%%%%%%%%%%%%%%%%%%

\section{Supplemental figures}
\label{app:A}

\begin{figure}[tb]
    \includegraphics[width=0.48\textwidth]{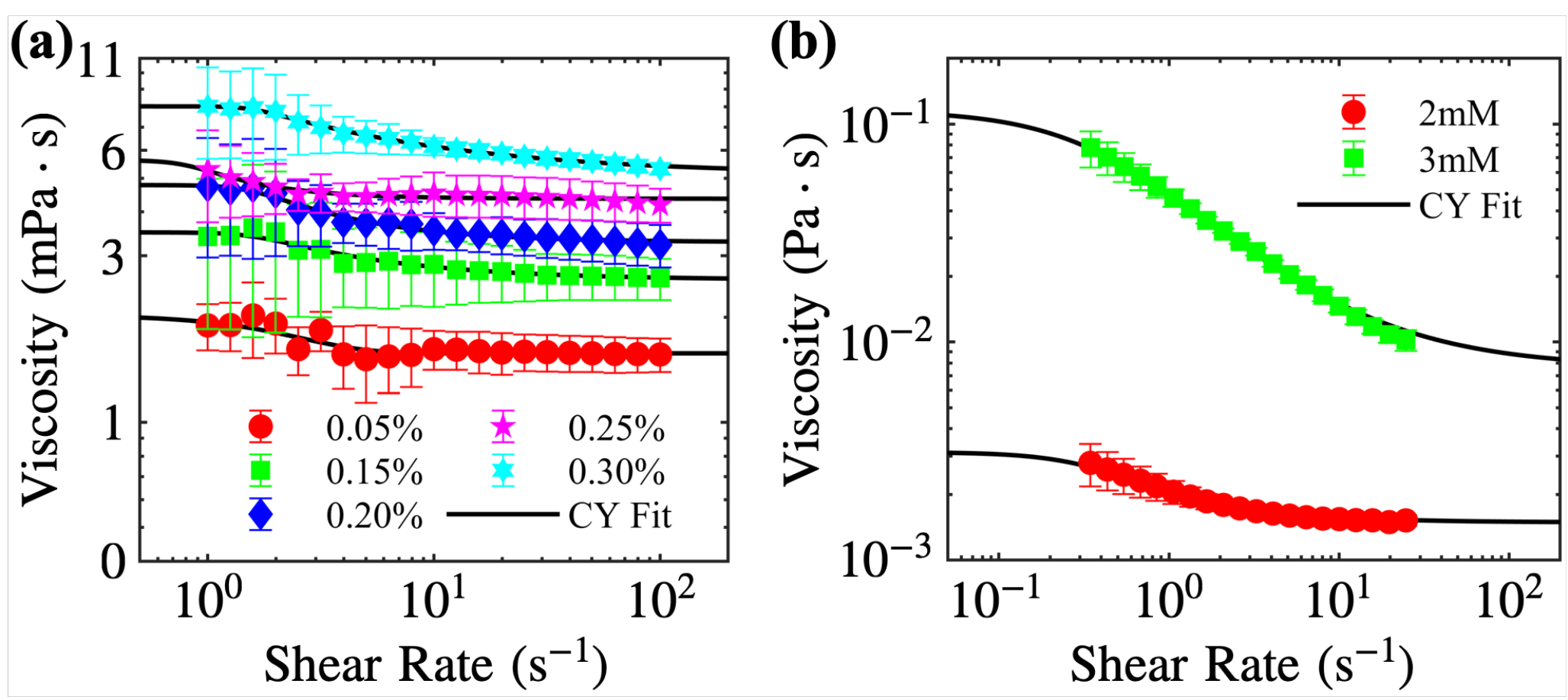}
    \caption{\label{fig:S1}
    Steady-state shear viscosity of (a) PAAM solutions and (b) CTAB/NaSal solutions at various concentrations. 
    }
\end{figure}

% Figure 5
Figure\,\ref{fig:S1}(a) and (b) show the steady-state shear viscosity of the PAAM solutions and the CTAB/NaSal solutions, respectively. 
Such data (points with error bars) were measured using a stress-controlled rheometer (TA Instruments Discovery Hybrid Rheometer DHR-2) equipped with a cone-plate geometry (40 mm diameter, 4$^\circ$ cone angle).
The black solid line denotes the fitting in terms of the Carreau-Yasuda mode.

%%%%%%%%%%%%%%%%%%%%%%%%%%%%%%%%%%%%%%%%%%%%%%%%%%%%%%%%%%%%%%%%%%%%%%%%%%%%

\begin{figure}[tb]
    \includegraphics[width=0.48\textwidth]{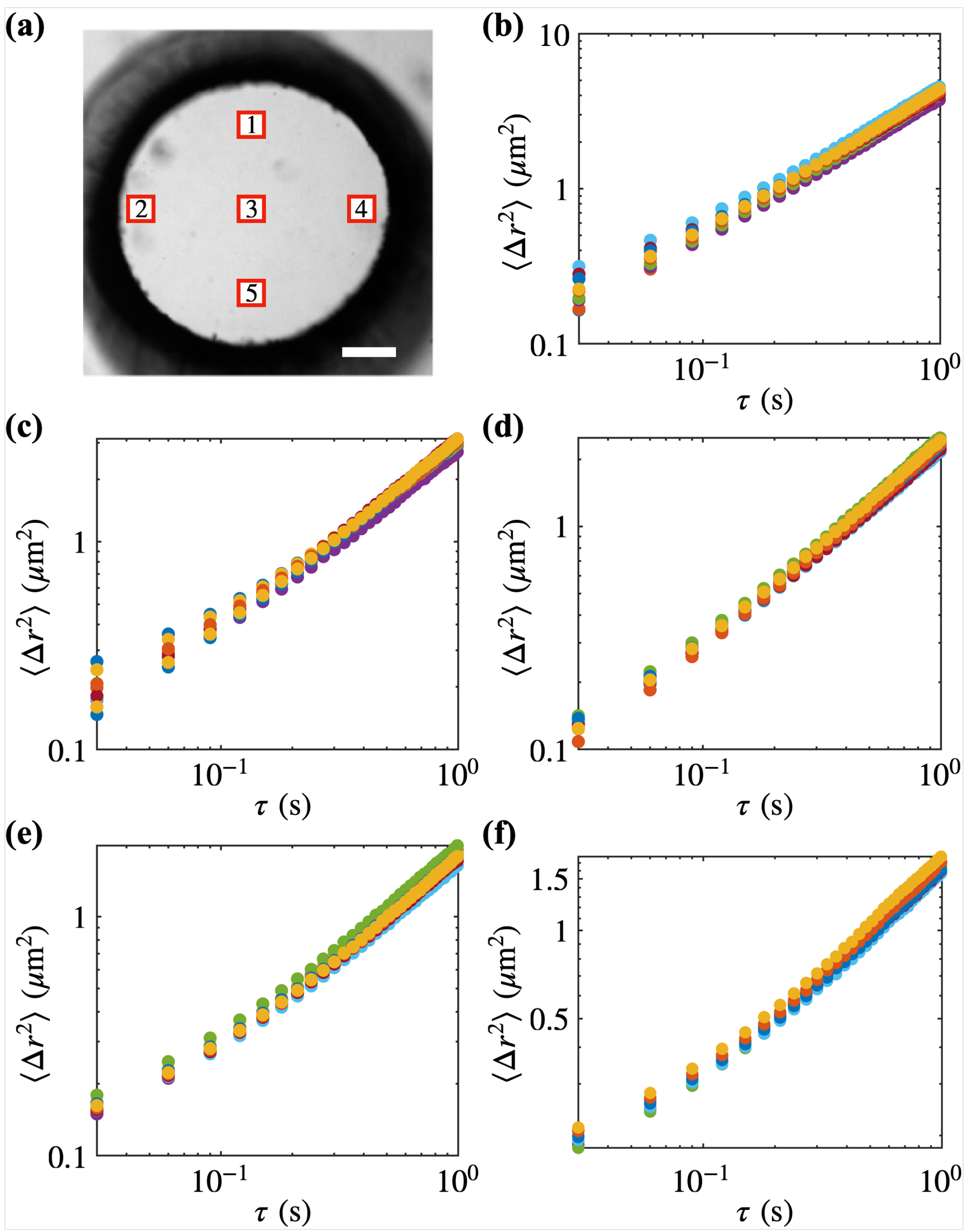}
    \caption{\label{fig:S2}
    (a) Sampling window (1 to 5) in a chamber with diameter equals 1 mm (scale 200 $\mu$m). 
    Time-dependent mean square displacements of fluorescent particles for various sampling windows and the PAAM solution of concentrations (b) 0.05\%, (c) 0.15\%, (d) 0.2\%, (e) 0.25\%, and (f) 0.3\%. 
    }
\end{figure}

Figure\,\ref{fig:S2}(a) shows the locations in the sample chamber for the microrheology measurement.
Figure\,\ref{fig:S2} (b) to (f) shows the representative time-dependent MSDs
in the PAAM solutions for various locations and the PAAM solutions in concentrations 0.05\%, 0.15\%, 0.2\%, 0.25\% and 0.3\%, respectively. 
Within each concentration, the curves nearly overlap, indicating the  homogeneous distribution of polymers. 

% Figure 6

%%%%%%%%%%%%%%%%%%%%%%%%%%%%%%%%%%%%%%%%%%%%%%%%%%%%%%%%%%%%%%%%%%%%%%%%%%%%

% Figure 7
\begin{figure}[tb]
    \includegraphics[width=0.48\textwidth]{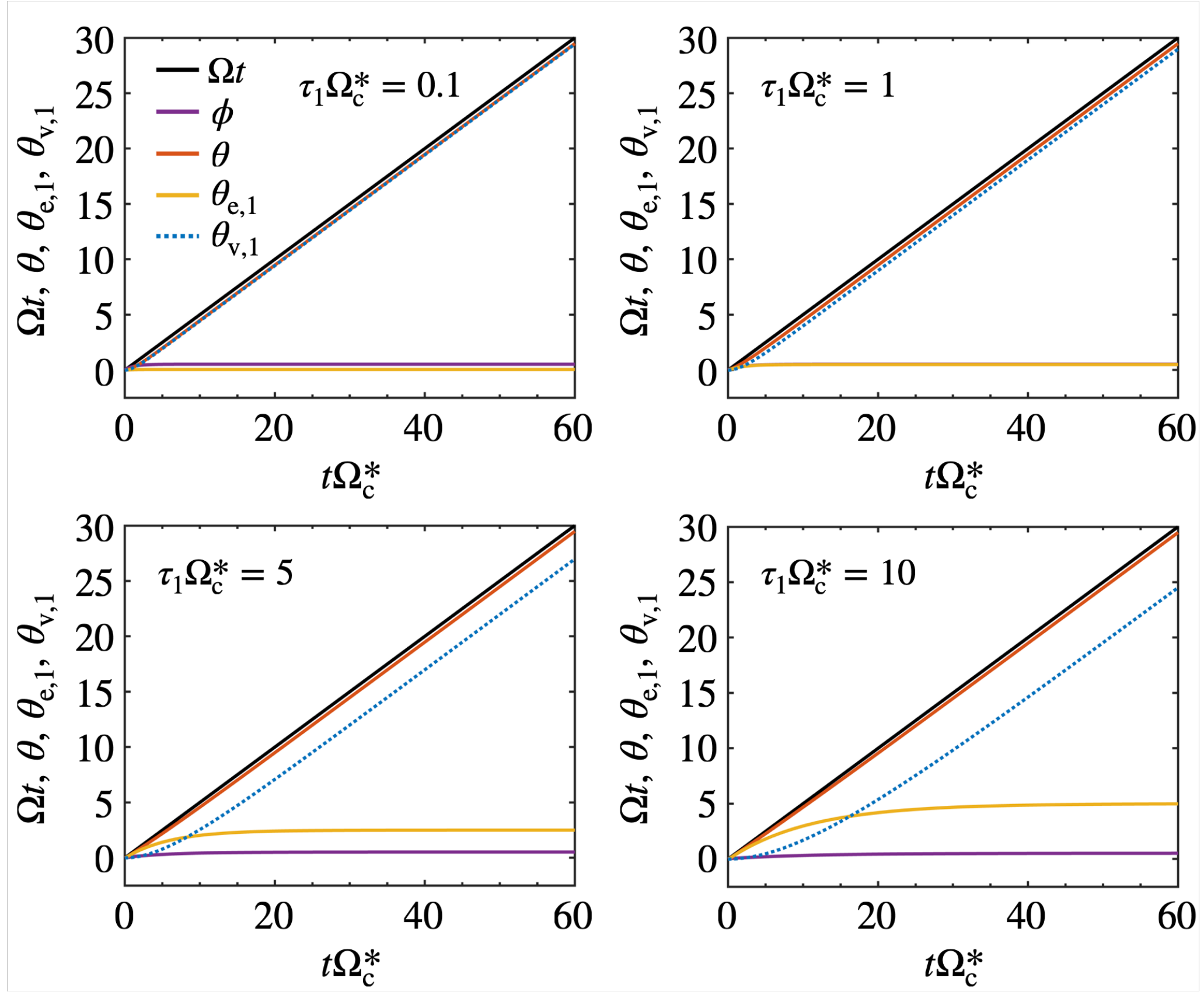}
    \caption{ \label{fig:S3}
    Time evolution of key angular variables for $\Omega / \Omega^\ast_\mathrm{c} = 0.5$ (in the synchronous regime), $\eta_\mathrm{p} / \eta_\mathrm{s} = 10$, and various values of $\tau \Omega^\ast_\mathrm{c}$.
    }
\end{figure}

Figure\,\ref{fig:S3} shows the time evolution of key angular variables for representative condition of $\eta_\mathrm{p} / \eta_\mathrm{s} = 10$ and $\Omega / \Omega^\ast_\mathrm{c} = 0.5$ (in the synchronous regime).
It is observed that with growing time, both curves of $\phi$ and $\theta_{\mathrm{e},1}$ approach plateaus, which correspond to steady states of the system. 
The non-zero steady value of $\theta_\mathrm{e,1}$ indicates finite storage deformation that cannot be relaxed.
Increasing $\tau \Omega^\ast_\mathrm{c}$, which is equivalent to increasing the relaxation time $\tau$ because of fixed $\Omega^\ast_\mathrm{c}$, results in elongated primary creep periods and the enhanced storage deformation during the steady-state creep.
However, the steady value of $\phi$ is changeless, because $\phi = \arcsin (\Omega / \Omega^\ast_\mathrm{c})$ is not a function of $\tau$.
It is noted that for the case of $\tau \Omega^\ast_\mathrm{c} = 1$, the curves of $\phi$ and $\theta_\mathrm{e,1}$ overlap.

%%%%%%%%%%%%%%%%%%%%%%%%%%%%%%%%%%%%%%%%%%%%%%%%%%%%%%%%%%%%%%%%%%%%%%%%%%%%

% Figure 8
\begin{figure}[tb]
    \includegraphics[width=0.48\textwidth]{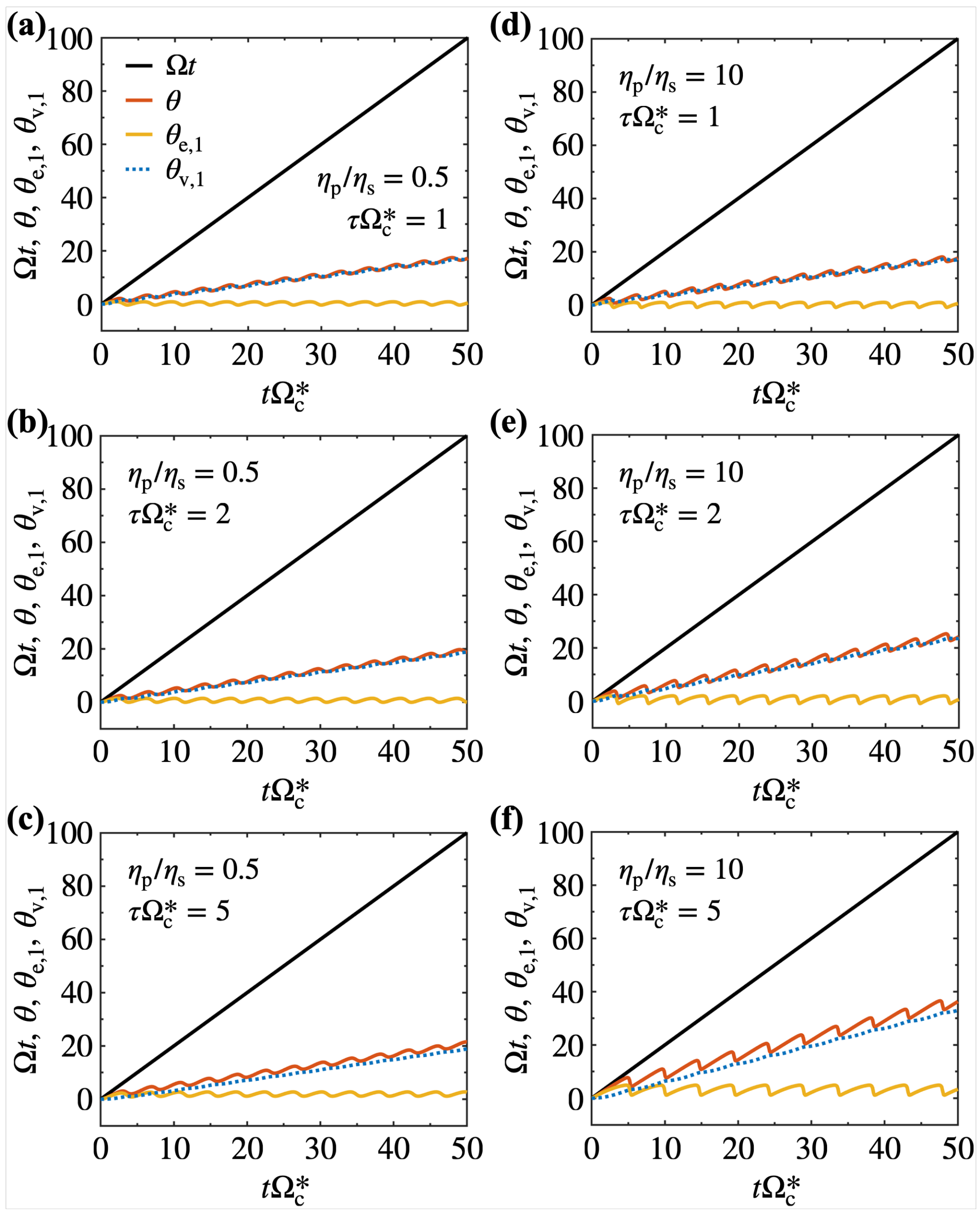}
    \caption{\label{fig:S4}
    Time evolution of key angular variables for $\Omega / \Omega^\ast_\mathrm{c} = 2$ (in the asynchronous regime), (a-c) $\eta_\mathrm{p} / \eta_\mathrm{s} = 0.1$ and (d-f) $\eta_\mathrm{p} / \eta_\mathrm{s} = 10$, and various values of $\tau \Omega^\ast_\mathrm{c}$.
    }
\end{figure}

Figure\,\ref{fig:S4} shows the time evolution of key angular variables for $\Omega / \Omega^\ast_\mathrm{c} = 2$ (in the asynchronous regime), and $\eta_\mathrm{p} / \eta_\mathrm{s} = 0.1$ (a to c) and $10$ (d to f).
It is observed that for the fixed $\eta_\mathrm{p} / \eta_\mathrm{s}$, increasing $\tau \Omega^\ast_\mathrm{c}$ or, equivalently, $\tau$ leads to a transition from viscous to elastic dominance in the sawtooth shape of the $\theta$ curve. 
Moreover, all the sawtooth amplitude, period, and the time-averaged value of $\theta_{\mathrm{e},1}$ increases as growing $\tau$.
This result indicates that for large relaxation time, the particle's `wobbling' becomes slow but significant in amplitude.  

%%%%%%%%%%%%%%%%%%%%%%%%%%%%%%%%%%%%%%%%%%%%%%%%%%%%%%%%%%%%%%%%%%%%%%%%%%%%

\section{Supplementary videos}
\label{app:B}

% Movie S1
Movie\,S1 shows the synchronous rotation of a ferromagnetic particle that is suspended in 0.2\% w/v PAAM solution and driven by a rotating magnetic field of intensity 0.71 mT and frequency 1 Hz. 
The movie is accelerated to 0.2$\times$ real-time speed.

% Movie S2
Movie\,S2 shows the `wobbling' rotation of a ferromagnetic particle that is suspended in 0.25\% w/v PAAM solution and driven by a rotating magnetic field of intensity 0.71 mT and frequency 2.5 Hz. 
The movie is accelerated to 0.3$\times$ real-time speed.

%%%%%%%%%%%%%%%%%%%%%%%%%%%%%%%%%%%%%%%%%%%%%%%%%%%%%%%%%%%%%%%%%%%%%%%%%%%%
%%%%%%%%%%%%%%%%%%%%%%%%%%%%%%%%%%%%%%%%%%%%%%%%%%%%%%%%%%%%%%%%%%%%%%%%%%%%

\bibliography{Main}% Produces the bibliography via BibTeX.
\end{document}